\documentclass[jkps,preprint,fleqn,showpacs,showkeys]{revtex4}
\usepackage[pdftex]{graphicx}
\usepackage{amssymb}
\usepackage{amsmath}
\usepackage{bm}
\usepackage{appendix}
\usepackage{footnote}
\begin{document}
\setcounter{page}{0}
\title[]{A Data-driven Event Generator for Hadron Colliders\\using  Wasserstein Generative Adversarial Network}
\author{Suyong \surname{Choi}}
\author{Jae Hoon \surname{Lim}}
\email{chaosbringer@korea.ac.kr}
\affiliation{Department of Physics, Korea University, Seoul 02841, Republic of Korea}
\begin{abstract}
Highly reliable Monte-Carlo event generators and detector simulation programs are important for the precision measurement in the high energy physics.
Huge amounts of computing resources are required to produce a sufficient number of simulated events.
Moreover, simulation parameters have to be fine-tuned to reproduce situations in the high energy particle interactions which is not trivial in some phase spaces in physics interests.
In this paper, we suggest a new method based on the Wasserstein Generative Adversarial Network (WGAN) that can learn the probability distribution of the real data.
Our method is capable of event generation at a very short computing time compared to the traditional MC generators.
The trained WGAN is able to reproduce the shape of the real data with high fidelity.
\end{abstract}


\maketitle
\thispagestyle{empty}
\section{INTRODUCTION}
There have been lots of important scientific discoveries at the Large Hadron Collider (LHC).
With the increasing integrated luminosity at the LHC, extremely rare and complicated physics processes could be studied such as the production of multiple massive quarks and vector bosons.
For the precise measurement of the physics involving these particles, we need a very good understanding in the background processes.
Due to the large theoretical uncertainties, it is recommended to estimate backgrounds from the real data.
In addition, MC simulations of the backgrounds are statistically limited in the phase space of interest.

The Generative Adversarial Network (GAN)~\cite{GAN} is a Deep Learning technique which can generate new fake image data.
The idea is applied to the detector simulation in the High Energy Physics (HEP), and CaloGAN~\cite{CALOGAN, 3DCAL} is one of the first applications. 
Then, the ideas on generating derived event variables have been also explored to study $Z\rightarrow\ell^+\ell^-$ or two jet events~\cite{GANZmm,GANDijet}.

In this paper, we applied one of GANs to produce kinetic event variables of HEP data. We explored Wasserstein GAN (WGAN) which is known to give improved results than the traditional GAN~\cite{WGAN, WGANimpr, WGANinHEP}.

\section{The Generative Adversarial Network}

A GAN is composed of two neural networks, ``generator" and ``discriminator".
The generator network, $G$, is supposed to create a sample of data from a set of variables called as latent variables.
By choosing a point in the latent variable space, it produces a data sample which is similar to the real data.
In other words, it can be considered as a multi-dimensional mapping between the latent space and the event data space.
The discriminator network $D$ distinguishes between the generated ``fake'' data and the ``real'' data that is used for training.
Two networks compete during the training, $G$ produces fake data as similar to the real data as possible and $D$ tries to discriminate generated data from the real data.
After successful training, any random point in the latent space can be mapped to the event data by the $G$ network.

The training strategy of the GAN is minimizing each loss function for the $D$ and $G$. 
The loss function of the discriminator network $L_{D}$ is defined by taking the average of the discriminator outputs over the fake data and the real data.
The loss function of the generator network $L_{G}$ is defined by the negative value of the average of the discriminator outputs over the fake data.
\begin{eqnarray}
L_{D} & = & \langle D\rangle_{fake} - \langle D\rangle_{real}\\
L_{G} & = & -\langle D\rangle_{fake}
\end{eqnarray}
When $L_D$ is minimized in a step, the discriminator network weights are updated so that the discriminator output for the real data would tend to 1 while for the fake data it would approach 0.
On the other hand, when $L_G$ is being minimized, the discriminator network is not modified, but the generator network weights are updated.

Special treatments are needed to overcome the issues of the traditional GAN.
First, it is known that the GAN tends to produce smoother distributions compared to the real data~\cite{NN, WGAN, GANZmm}.
Second, the landscape of the loss function of GAN is well known to be in a saddle point, drift or show sudden jumps, therefore it has stability problems.
In WGAN, a penalty term related with $\langle \nabla D\rangle_{P}$ is added to the $L_D$ to penalize the gradients, where $P$ is a point between the real data and the fake data in the event data space.
So the updates to the network weights and biases become more gradual~\cite{WGAN, WGANimpr}.
This assures convergence of the network parameters and stability of the training.

\section{Application of WGAN in $\bm{pp \rightarrow b\bar{b}\gamma\gamma}$ production}
A HEP data can be considered as a set of numbers.
For example, HEP data are kinematic variables of final state physics objects such as hadronic jets, leptons, or the missing transverse momentum.
If the WGAN is applied to this case, a point in the latent space will be mapped to an event containing particle momenta components $(p_x, p_y, p_z)$. 

In this study, we focused on $pp \rightarrow b \bar{b}\gamma\gamma$ process, which is one of the backgrounds to search for the double Higgs boson production at the LHC.
These events would allow us to probe the self-coupling of the Higgs and reconstruct the Higgs potential.
One of the important search modes of the double Higgs production proceeds through $pp\rightarrow HH\rightarrow b\bar{b}+\gamma\gamma $. 

As an illustration of the WGAN method, we try to mimic a Monte Carlo simulated data, a proxy for the real data.
We generated 1 million events of non-resonant production of $pp\rightarrow b\bar{b}\gamma\gamma$ at $\sqrt{s}=14$ TeV at the leading order with \textsc{Madgraph 5}~\cite{MG5}.
\textsc{Pythia 8} was used to simulate subsequent particle shower and hadronization process~\cite{PYTHIA}.
A fast parametrized detector simulation and reconstruction were performed with \textsc{Delphes 3} software package using the default CMS~\cite{Delphes3,CMS} detector settings without pileup (multiple simultaneous $pp$ interactions) effects.
Generated events are required to satisfy selection criteria:
\begin{itemize}
\item two photons with the transverse momentum $p_T >10\ {\rm GeV}$ and the pseudorapidity $|\eta|<2.4$.
\item at least two hadronic jets reconstructed with anti-$k_T$ algorithm~\cite{antiKT} with 0.4 angular distance ($\Delta R = \sqrt{\Delta\eta^2+\Delta\phi^2}$, where $\phi$ is the azimuthal angle) with $p_T>20\ {\rm GeV}$ and $|\eta|<2.4$.
\item at least two tagged $b$-jets.
\end{itemize}
After the event selection, about 70,000 events remained and are used for the training.

For the implementation of the WGAN, we use \textsc{TensorFlow 1.8.0}~\cite{TensorFlow} with \textsc{Keras 2.2.0}~\cite{KERAS}.
We build a modified WGAN composed of one generator network and 10 discriminator networks.
To construct the generator network and the discriminator networks, we apply the multilayer perceptron model.
Each network has four hidden layers with different node sizes of 2048, 2048, 1024, and 256 with the Rectified Linear Unit as the activation function of the neurons.
Between the layers, we set a dropout rate of 0.05.
The generator network is configured to accept input from a 22-dimensional latent space.
For the discriminator network, we choose the softmax function as an activation function at the output layer.

\begin{figure}
\includegraphics[width=13cm]{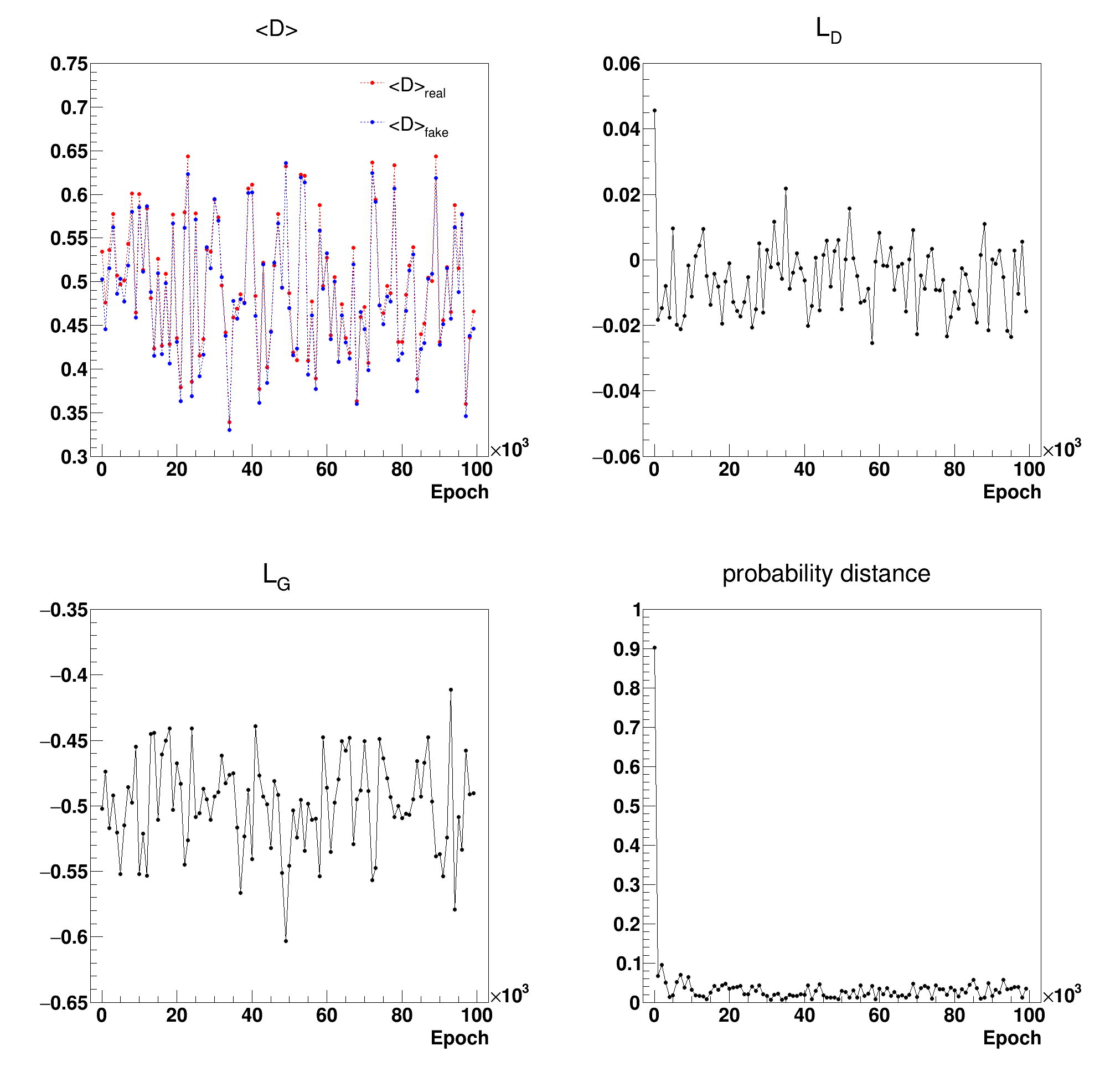}
\caption{\label{fig:train}
Training history of $\langle D\rangle$, $L_{D}$, $L_{G}$, and the probability distance.
}
\end{figure}

The WGAN was trained adversarially on about 70,000 events with a batch size of 256 for 100,000 epochs.
Momentum components $(p_x, p_y, p_z)$ of the leading two photons and two leading $b$-jets are used as inputs into the WGAN.
Particle momenta are scaled down for the stability of the training.
Training is done by Adam with a learning rate of $5\times 10^{-5}$.
Training takes 20 hours with NVIDIA Tesla P100 graphic processor unit and the event generation with the trained WGAN just takes a few seconds for 10 million events. 

Generated events by WGAN were not uniform in the azimuth or symmetric in pseudorapidity.
These symptoms seemed to stem from using a single discriminator in the adversarial training.
Interestingly, when we create many discriminators, we observe that distributions recover the expected symmetry in angular distributions.
By employing different discriminators, the generator should learn to cover the whole phase space.
We employ multiple discriminator networks to train the generator~\cite{GMAN}, and each discriminator network $D^{\,i}$ is trained to minimize own $L_{D}^{\,\,i}$ separately.

The probability distance is defined as the total variation distance between the discriminator output distribution of input real events and that of generated fake events. 
As shown in Fig.~\ref{fig:train}, the probability distance is minimized to 0.05, and $\langle D\rangle$, $L_{D}$, and $L_{G}$ show adversarial training in progress for 100,000 epochs.

\begin{figure}
\includegraphics[width=17cm]{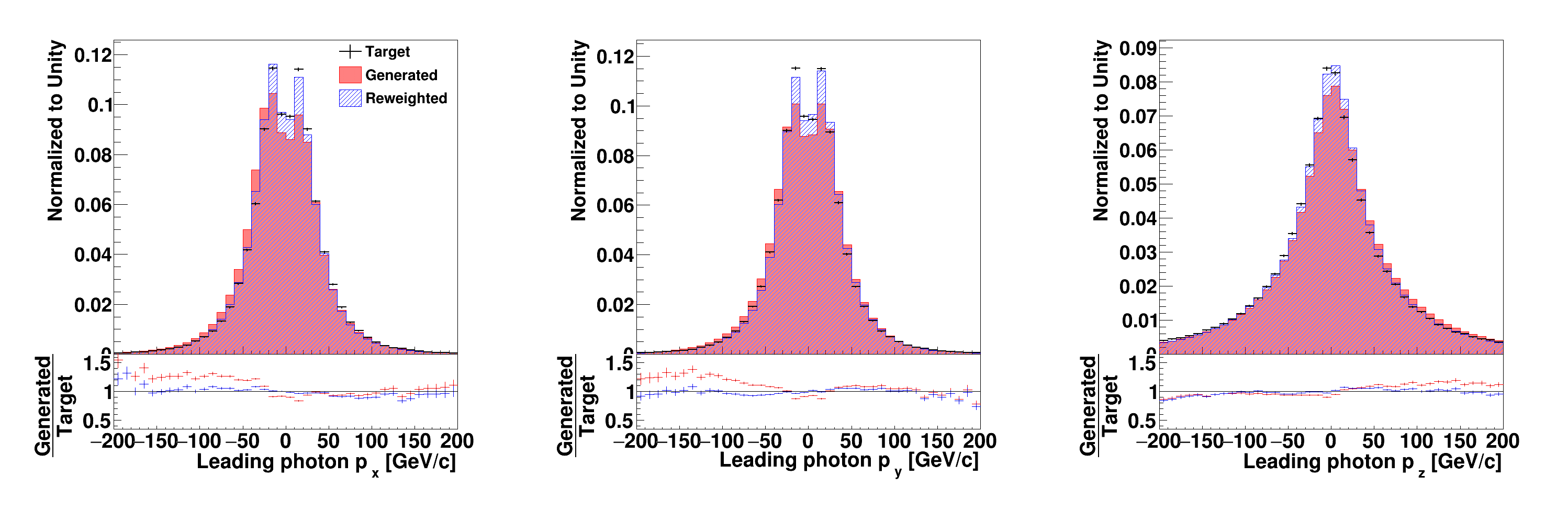}
\caption{\label{fig:input}
The distributions of three momentum components $(p_x, p_y, p_z)$ of the leading photon. Black lines represent the target events (real data), red distributions represent the generated events (fake data before reweighting), and blue distributions represent the reweighted events (fake data after reweighting).}
\end{figure}

\begin{figure}
\includegraphics[width=13cm]{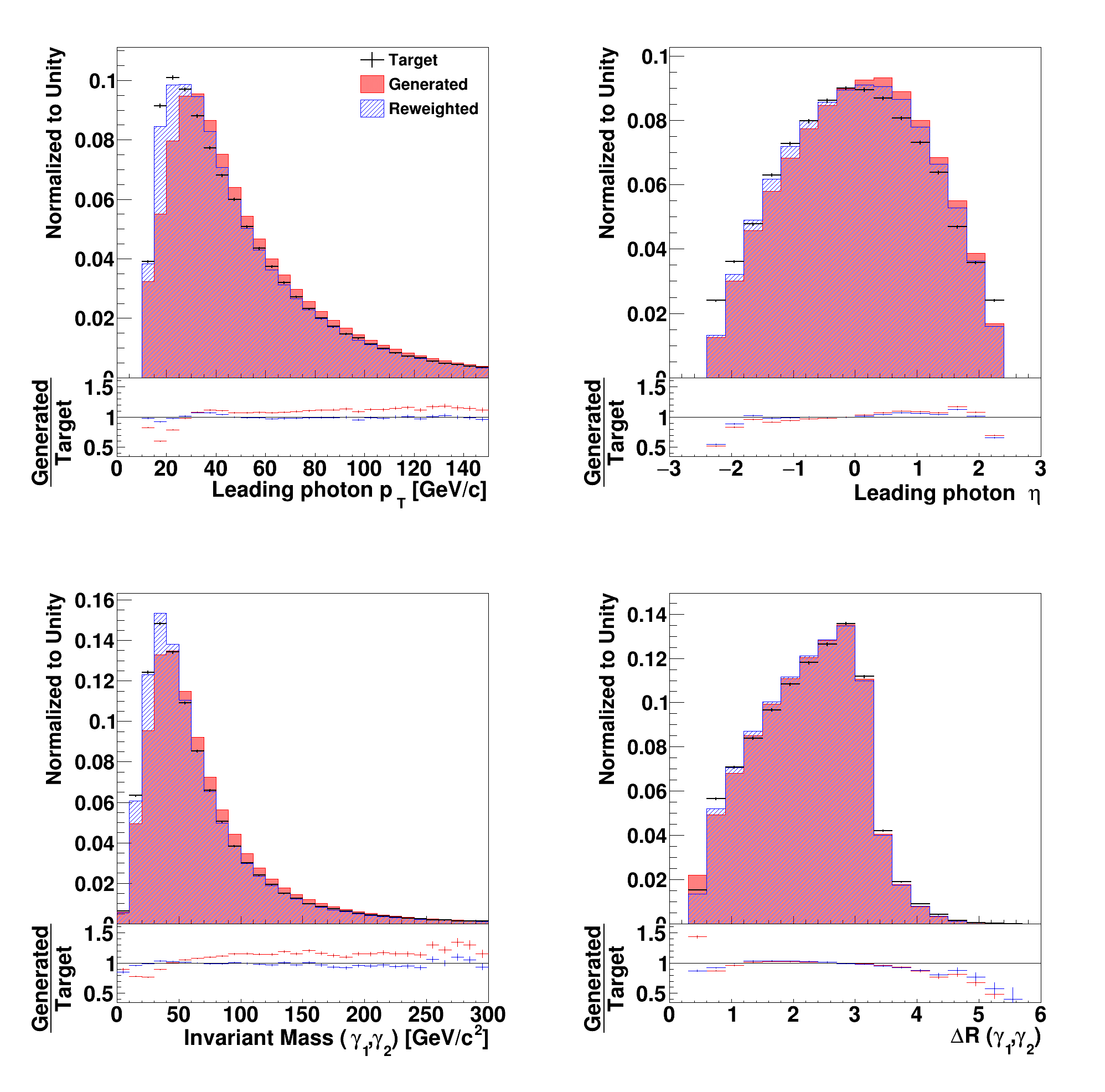}
\caption{\label{fig:output}
The distributions of  $p_{T}$ and $\eta$ of the leading photon, the invariant mass of two photons, and $\Delta R (\gamma_{1}, \gamma_{2})$. Black lines represent the target events (real data), red distributions represent the generated events (fake data before reweighting), and blue distributions represent the reweighted events (fake data after reweighting).
}
\end{figure}

The distributions of momentum components $(p_x, p_y, p_z)$ of the leading photon are in Fig. \ref{fig:input}, and the distributions of $p_T$, $\eta$, the invariant mass of two leading photons ($\gamma_1, \gamma_2$), and $\Delta R (\gamma_1, \gamma_2$) are in Fig. \ref{fig:output}. In each plot, the black line represents targeted events, the area filled with red color represents generated events by WGAN.
We observe a discrepancy between distributions of generated events and those of the targeted events.
Discrepancies stem from the characteristics of WGAN, which has a tendency to make smoother distributions than those of inputs.

To recover discrepancies, we derive weighting factors by investigating the difference between generated events and input data.
We used 24 variables - three momentum components of two leading photons and $b$-jets, the $p_T$ and the $\eta$ of two leading photons and $b$-jets, the invariant mass of two leading photons and two leading $b$-jets, $\Delta R (\gamma_1, \gamma_2$), and $\Delta R (b\textrm{-jet}_1, b\textrm{-jet}_2$).
We use a Gradient Boosted Decision Tree (GBDT)~\cite{GBDT} to derive weighting factors of each generated event.
40 trees with the maximum depth of 3 and the minimum number of events in a leaf of 200 are used for the GBDT.
In Fig. \ref{fig:input} and \ref{fig:output}, the reweighted distributions (blue color) have better agreements with input data.

\begin{figure}
\includegraphics[width=13cm]{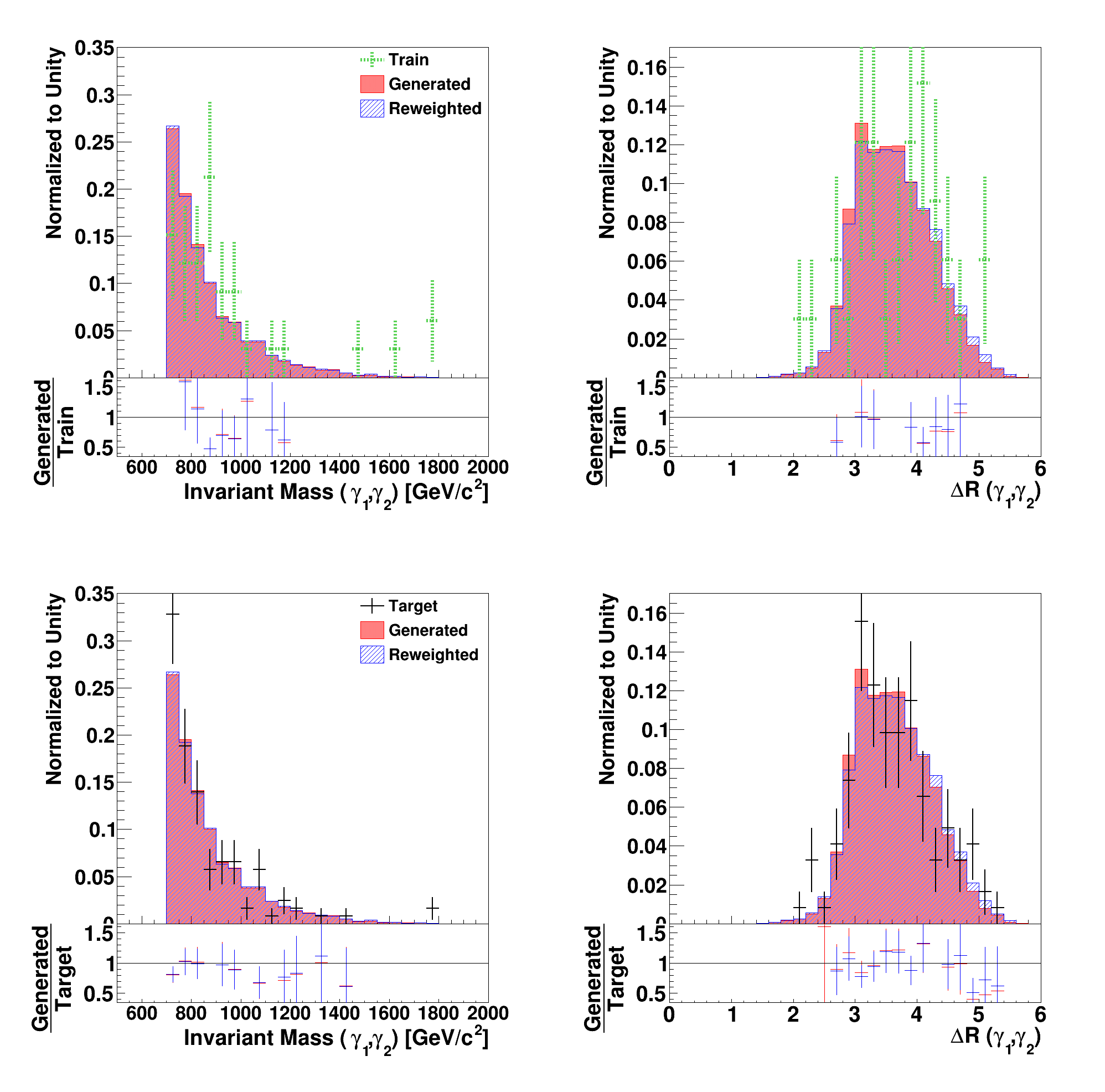}
\caption{\label{fig:small}
The distributions of the invariant mass of two photons and $\Delta R (\gamma_{1}, \gamma_{2})$.
The green dashed lines represent the training events (real data), the black lines represent the target events (real data), red distributions represent the generated events (fake data before reweighting), and blue distributions represent the reweighted events (fake data after reweighting).
}
\end{figure}

Figure \ref{fig:small} shows the invariant mass and $\Delta R$ distributions of two leading photons in the high mass region ($700 < M < 1800$ GeV/c$^2$).
The green dashed lines represent the training events, which have 4 times lower statistics than the targeted distributions and the black solid lines represent the targeted events. 
In this region, only 33 events of input data were used for the training, but the generated outputs show good agreement with the targeted distribution. 
This shows that the WGAN was able to reproduce the probability distributions in the this region even with the limited statistics.
It also indicates that the WGAN based algorithm can be considered as an alternative, data-driven event generator.
\section{CONCLUSIONS}
We developed a WGAN for reproducing $pp\rightarrow b\bar{b}\gamma\gamma$ events, one of the important background samples for double Higgs boson studies at the LHC.
The trained WGAN can generate events at a very short computing time compared to the traditional MC generators and reproduce the shape of the real data with high fidelity even with the limited statistics.
We expect that WGAN is a fast and faithful data-driven method for processes that are difficult to simulate in high energy physics. 
\begin{acknowledgments}
This work is supported by the National Research Foundation of Korea (NRF) under Contract No.NRF-2018R1A2B6005043, NRF-2020R1A2C3009918, and the BK21 FOUR program at Korea University, \textit{Initiative for science frontiers on upcoming challenges}.
\end{acknowledgments}

\newpage
\begin{appendices}

\section{Normalized distributions of momentum components}
\begin{figure}[h]
\includegraphics[width=17cm]{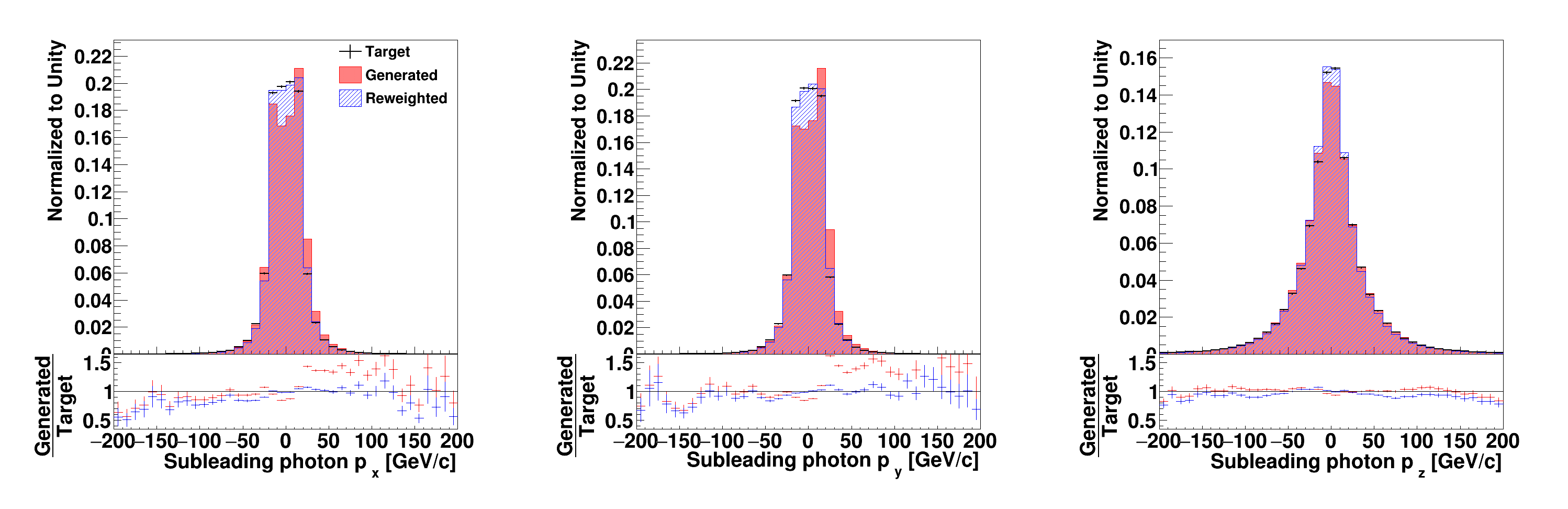}
\caption{
The distributions of three momentum components $(p_x, p_y, p_z)$ of the subleading photon.
}
\includegraphics[width=17cm]{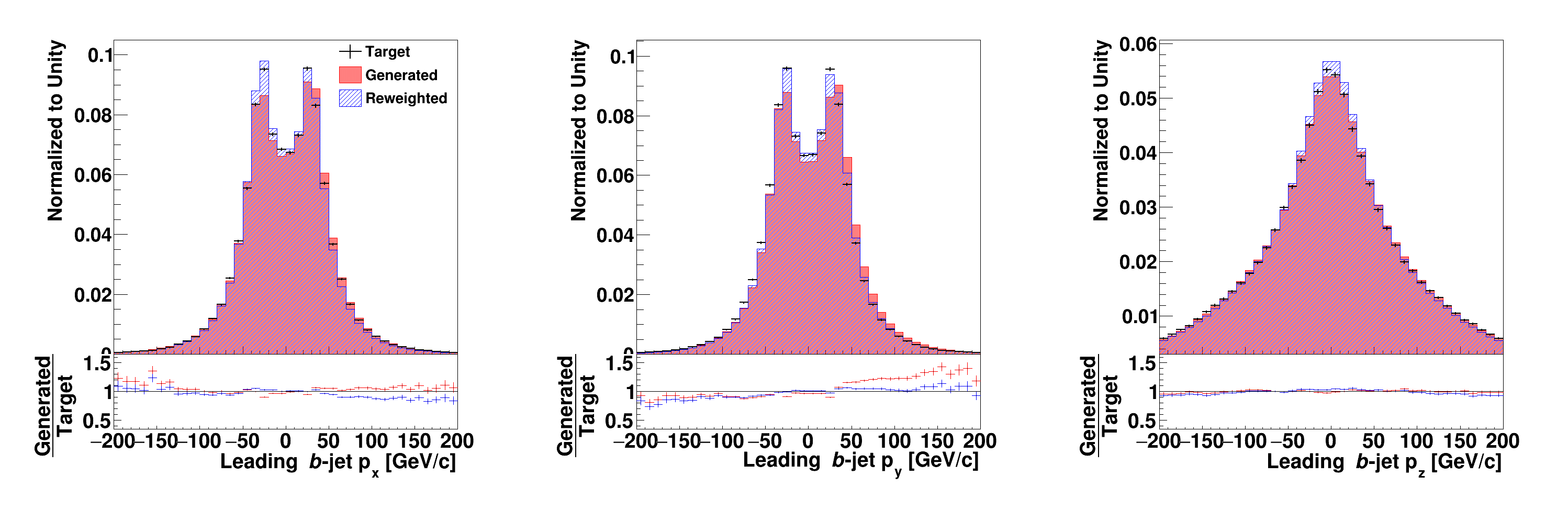}
\caption{
The distributions of three momentum components $(p_x, p_y, p_z)$ of the leading $b$-jet.
}
\includegraphics[width=17cm]{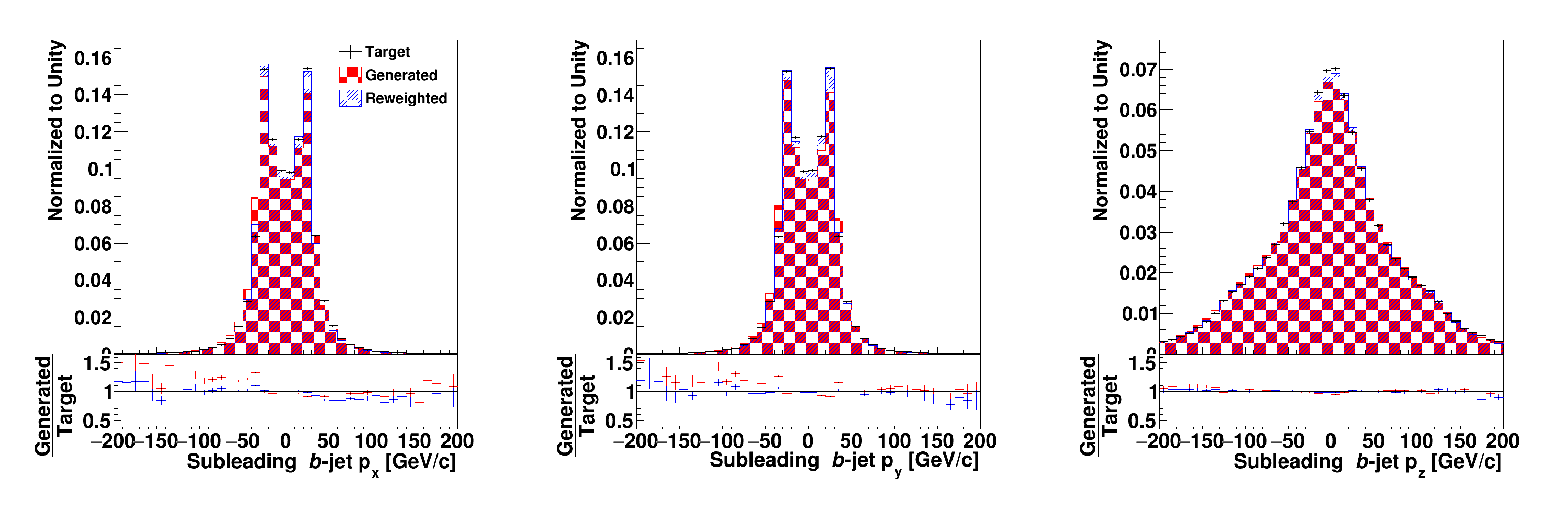}
\caption{
The distributions of three momentum components $(p_x, p_y, p_z)$ of the subleading $b$-jet.
}
\end{figure}
\newpage
\section{Normalized distributions of additional variables}
\begin{figure}[h]
\includegraphics[width=13cm]{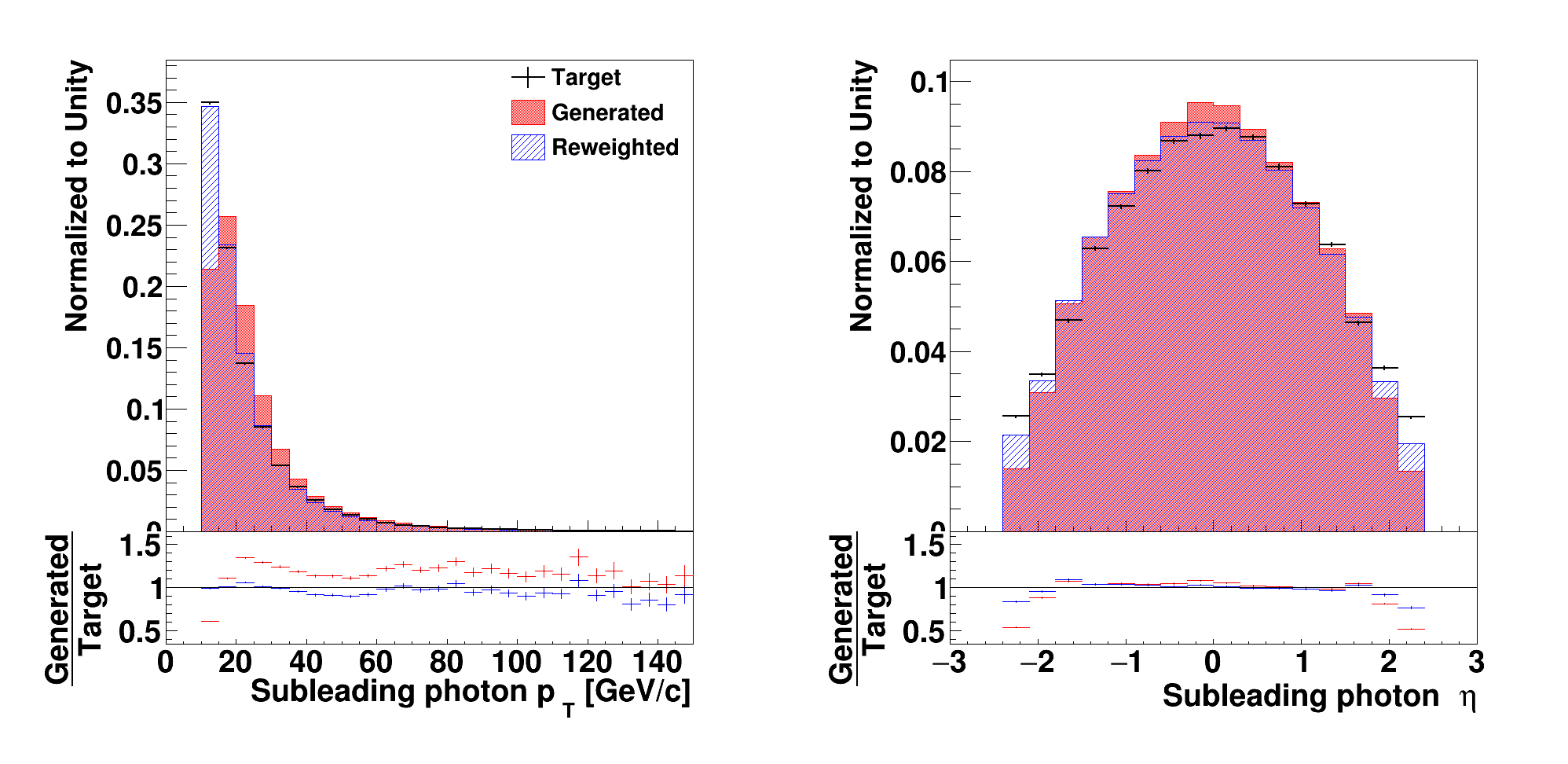}
\caption{
The distributions of  $p_{T}$ and $\eta$ of the subleading photon.}
\end{figure}
\begin{figure}[h]
\includegraphics[width=13cm]{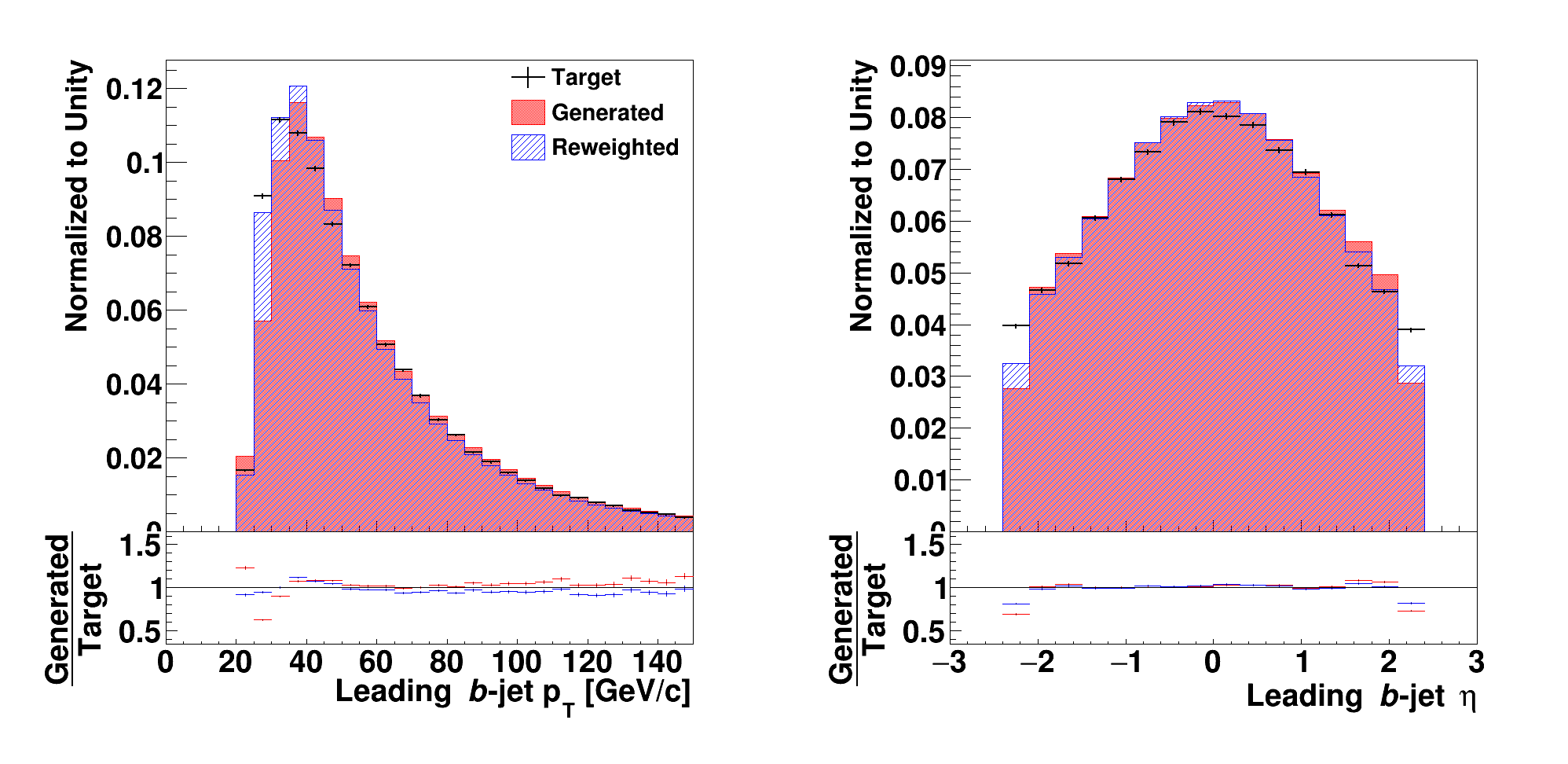}
\caption{
The distributions of  $p_{T}$ and $\eta$ of the leading $b$-jet.
}
\end{figure}
\begin{figure}[h]
\includegraphics[width=13cm]{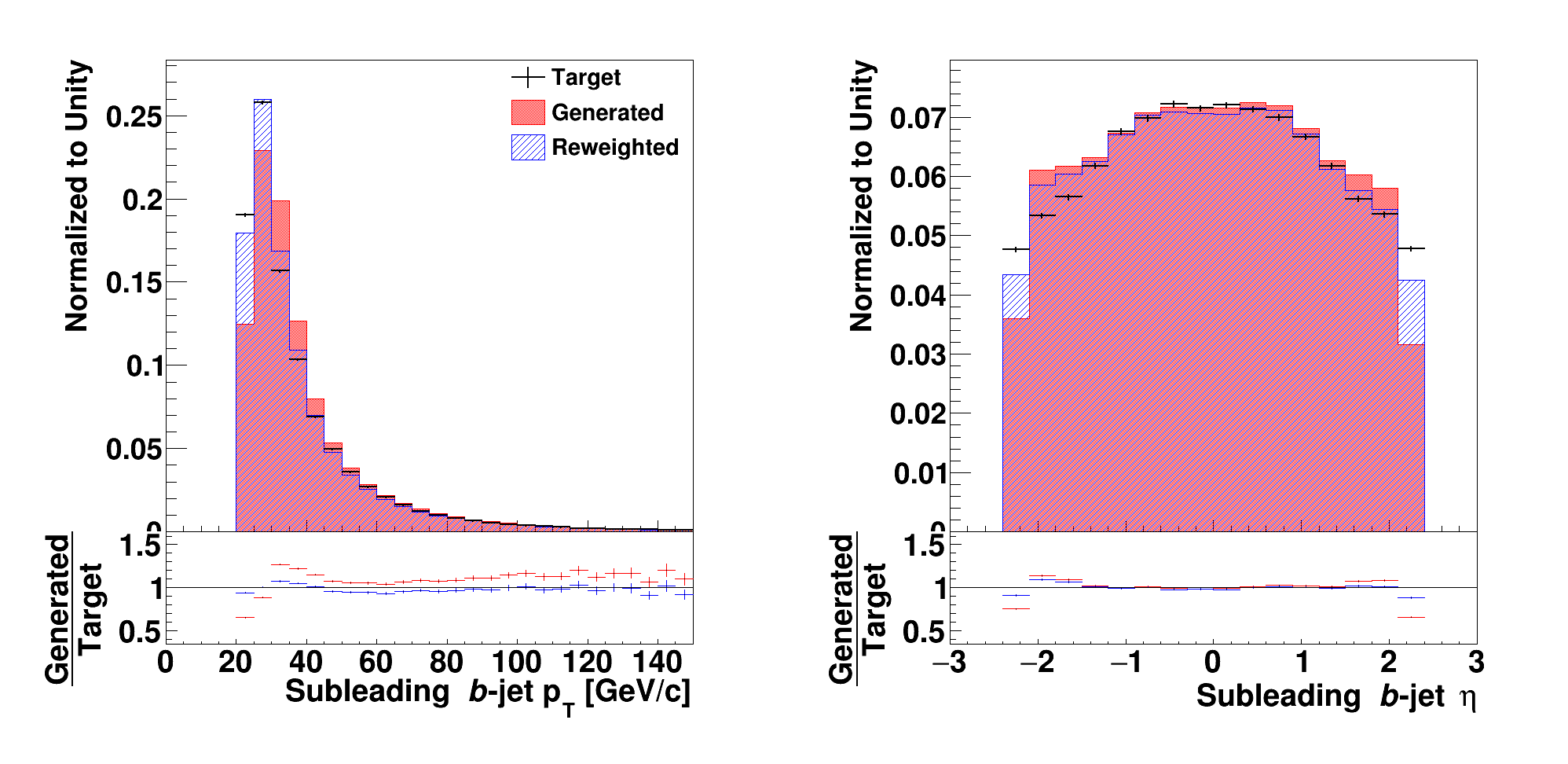}
\caption{
The distributions of  $p_{T}$ and $\eta$ of the subleading $b$-jet.
}
\end{figure}
\begin{figure}[h]
\includegraphics[width=13cm]{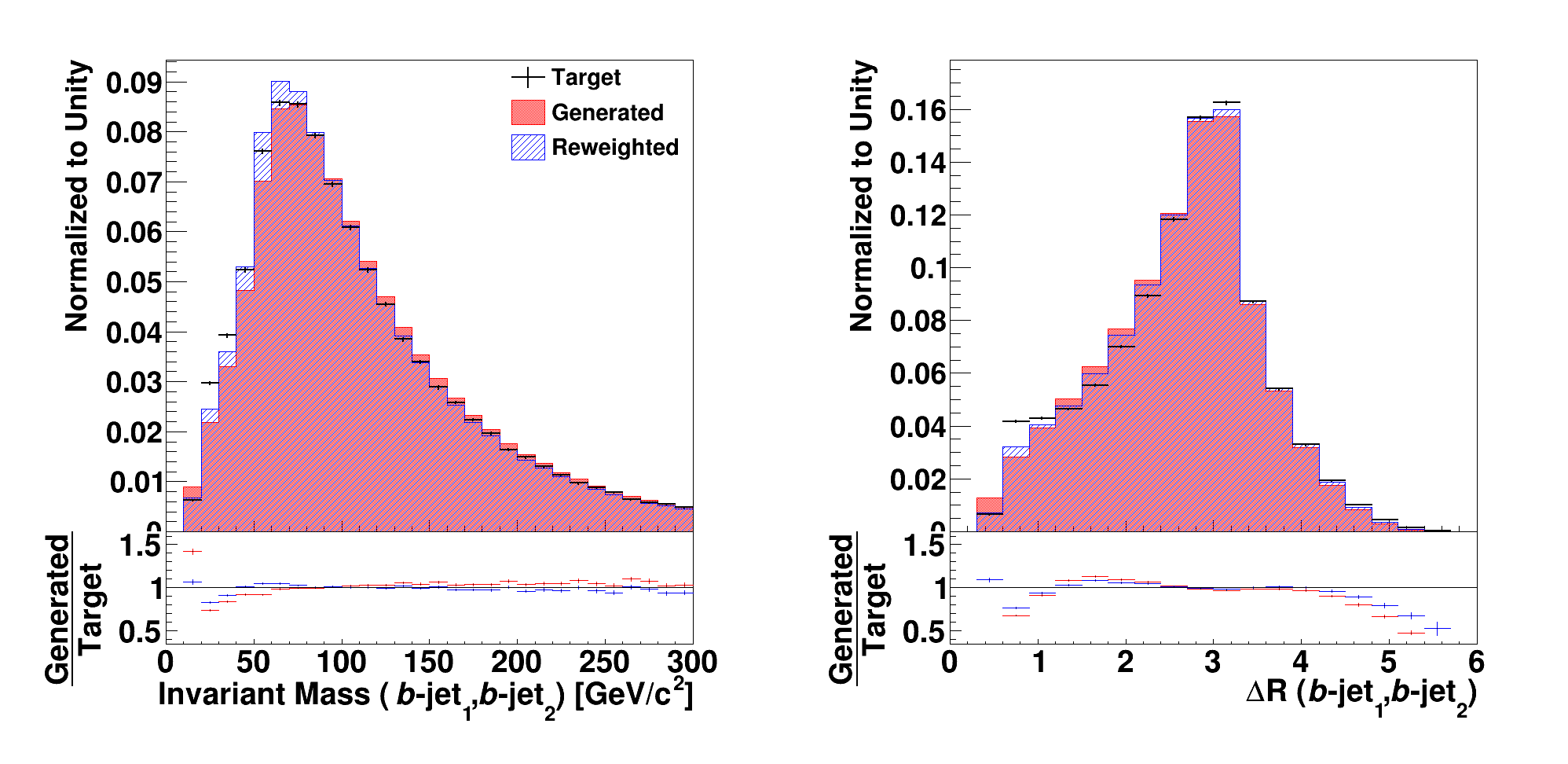}
\caption{
The distributions of invariant mass of two $b$-jets and $\Delta R (b\textrm{-jet}_1, b\textrm{-jet}_2$).
}
\end{figure}
~\\
\vspace{200px}
~\\
\newpage
\section{Normalized distributions of the control region}
\begin{figure}[h]
\includegraphics[width=13cm]{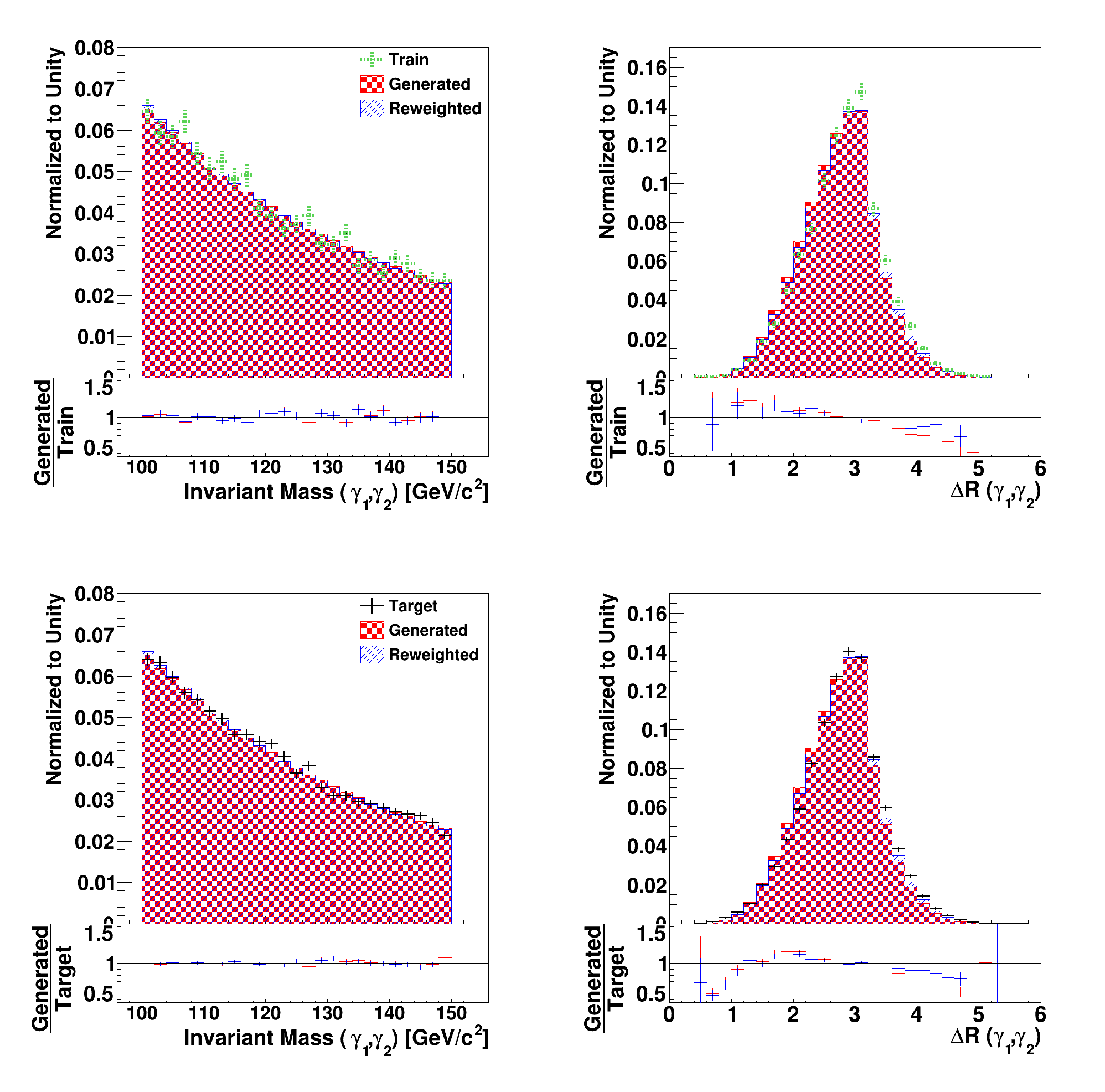}
\caption{
The distributions of the invariant mass of two photons and $\Delta R (\gamma_{1}, \gamma_{2})$ with the condition of [100, 150] GeV.
}
\end{figure}

\end{appendices}

\end{document}